\documentclass[letterpaper,10pt]{article} 

\usepackage{osameet3} 

\usepackage{amsmath,amssymb}
\usepackage{float}
\usepackage[colorlinks=true,bookmarks=false,citecolor=blue,urlcolor=blue]{hyperref} 
\usepackage{blindtext, graphicx, tabularx, multicol, lipsum, eqnarray,amsmath,chemformula,multirow,threeparttable, subfigure}
\usepackage{algorithm}
\usepackage[noend]{algpseudocode}
\usepackage{caption}

\begin{document}

\title{Optimizing Coherent Integrated Photonic Neural Networks under Random Uncertainties}

\author{Sanmitra Banerjee\textsuperscript{1}, Mahdi Nikdast\textsuperscript{2}, and Krishnendu Chakrabarty\textsuperscript{1}}
\address{\textsuperscript{1}ECE Dep., Duke University, Durham, NC, USA \textsuperscript{2}ECE Dep., Colorado State University, Fort Collins, CO, USA}
\email{sanmitra.banerjee@duke.edu, mahdi.nikdast@colostate.edu, krish@duke.edu}

\copyrightyear{2021}

\begin{abstract}
We propose an optimization method to improve power efficiency and robustness in silicon-photonic-based coherent integrated photonic neural networks. Our method reduces the network power consumption by 15.3\% and the accuracy loss under uncertainties by 16.1\%. 
\end{abstract}\vspace{-0.04in}

\section{Introduction}
\begin{figure}[H]
  \centering
  \subfigure[An example of a 4$\times$4 linear layer in an IPNN]{
\includegraphics[width=.45\textwidth]{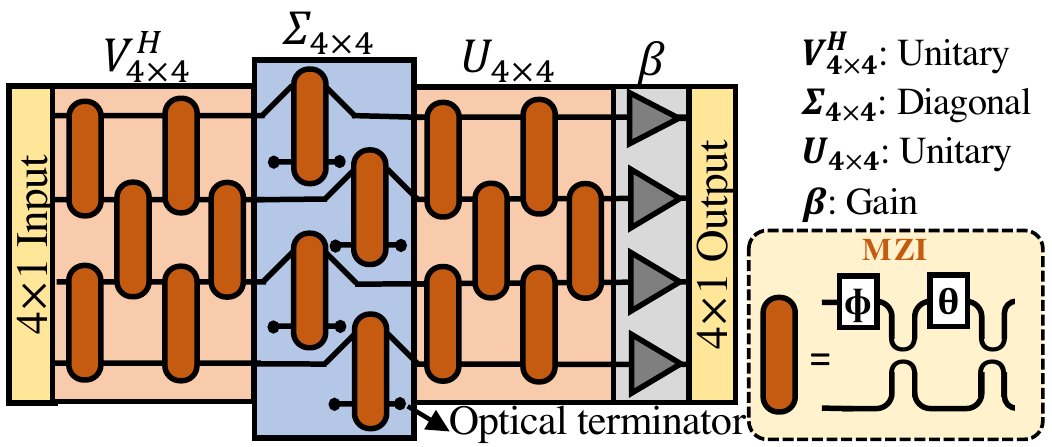}
}\hspace{-0.08in}%
\subfigure[MZI transfer matrix deviation]{
\includegraphics[width=.26\textwidth]{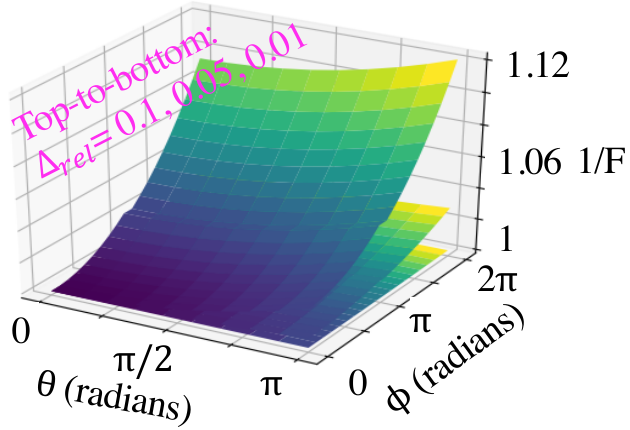}
}\hspace{-0.02in}%
\subfigure[Inferencing-accuracy loss]{
\includegraphics[width=.26\textwidth]{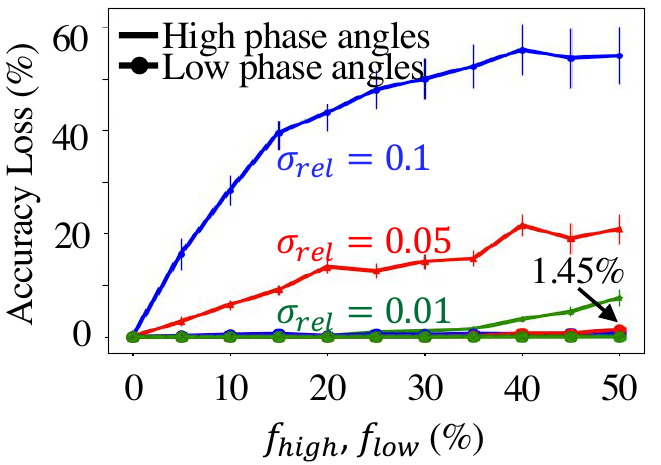}
}
\vspace{-1em}
  \caption{(a) Linear-layer representation using MZI arrays. (b) Deviation in the transfer matrix of an MZI from its nominal value when different variations ($\Delta_{rel}$) are introduced in $\theta$ and $\phi$. (c) Phase angles in each layer are ranked in descending order. For each $f_{high}$ and $f_{low}$, the uncertainty in each phase angle is sampled from a Gaussian distribution with its nominal value as mean ($\mu$) and standard deviation $\sigma=\sigma_{rel}\cdot \mu$. The plot (whiskers) show the mean (standard deviation) of the accuracy loss over 10 iterations. The nominal classification accuracy is 87.48\%.}
  \vspace{-1.5em}
  \label{fig1}
\end{figure}

Coherent integrated photonic neural networks (IPNNs) based on silicon photonics (SiPh) offer small-footprint, cost-effective, and energy-efficient optical matrix-vector multipliers (OMMs) with a computation time of only $O(1)$ in multi-layer perceptrons (MLPs). Using singular value decomposition (SVD), a matrix can be factorized into one diagonal and two unitary matrices; each of these can be realized using an array of Mach--Zehnder interferometers (MZIs)~\cite{clements2016optimal}, as shown in Fig. \ref{fig1}(a). Moreover, neural network training algorithms can be used to train the matrix parameters for different layers, each of which contains a set of weights that can be decomposed into different phase settings on each SiPh device (i.e., MZIs in Fig. \ref{fig1}(a)). As a result, any deviations in such adjusted phase settings (i.e., weights) will lead to network inferencing-accuracy losses. Unfortunately, the underlying SiPh devices in IPNNs are sensitive to inevitable fabrication-process variations (FPVs) and on-chip thermal crosstalk~\cite{cheng2020silicon}, both of which are a significant source of phase errors in IPNNs, and can cause up to 70\% loss in IPNN inferencing-accuracy~\cite{banerjee2020modeling}. Leveraging the non-uniqueness of SVD under reflections, we propose a novel optimization method to improve power efficiency and robustness in SiPh-based coherent IPNNs under random uncertainties---stemming from FPVs and thermal crosstalk---without affecting the inferencing-accuracy. In particular, we show that MZIs with higher adjusted phase angles are more susceptible to uncertainties. Accordingly, we minimize the phase angles in such MZIs to not only save tuning power but also improve network robustness. Our results based on simulating a fully-connected IPNN with 1374 tunable thermal phase shifters show up to 15.3\% and 16.1\% improvement in, respectively, the network power efficiency and robustness under uncertainties.\vspace{-0.1in}

\section{Theory: Impact of Uncertainties in SiPh-based Coherent IPNNs}
The transfer function of an MZI is given by $T=\left(\begin{smallmatrix}e^{i\phi}(e^{i\theta}-1)/2 && i(e^{i\theta}+1)/2\\ie^{i\phi}(e^{i\theta}+1)/2 && -(e^{i\theta}-1)/2 \end{smallmatrix}\right)$, where $\theta$ and $\phi$ are the phase angles (see Fig.~\ref{fig1}(a)). Under uncertainties, the deviated phase angles are $\tilde{\theta}=\theta(1+\Delta_{rel})$ and $\tilde{\phi}=\phi(1+\Delta_{rel})$, where $\Delta_{rel}$ denotes the relative change in the phase angles. To measure the ``closeness" between the deviated transfer matrix $\tilde{T}$ and $T$, fidelity  is defined as $F(T, \tilde{T})=\left|Trace(\tilde{T}^{\dagger}T)/N\right|^2$, where $F(T, \tilde{T})=$~1 if and only if $T=\tilde{T}$~\cite{walls2007quantum}. When $F$ decreases, the similarity between $T$ and $\tilde{T}$ decreases. Using this metric, Fig. \ref{fig1}(b) shows how $F$ changes for an MZI transfer matrix considering different values of $\Delta_{rel}$ (10\%, 5\%, and 1\%). As can be seen, an MZI with higher phase angles ($\theta$ and $\phi$ on x- and y-axis) is more susceptible to variations in $\theta$ and $\phi$ (note that the z-axis in Fig.~\ref{fig1}(b) denotes $1/F$). To verify this at the system-level, we consider a four-layer IPNN and observe its performance on the MNIST dataset. Each real-valued image in the MNIST dataset is converted to a complex feature vector of length 16 using a method based on fast Fourier transform~\cite{banerjee2020modeling}. A fully-connected network with two hidden layers of 16-complex valued neurons is implemented using the Clements design~\cite{clements2016optimal}. During inferencing, the adjusted phase angles in MZIs in each layer are ranked in a descending order, and then uncertainties---sampled from a Gaussian distribution---are introduced to the the top $f_{high}$\% and bottom $f_{low}$\% ranked phase angles in each layer. As Fig.~\ref{fig1}(c) shows, the network accuracy loss is catastrophic (up to $\approx$~60\%) when MZIs with higher phase angles are affected. In other words, the resilience of an IPNN against random uncertainties could be improved by minimizing adjusted phase angles in MZIs. Note that larger phase angles necessitate higher tuning-power consumption as the phase shift and tuning power are linearly proportional~\cite{jacques2019optimization}. Therefore, we propose a novel optimization method that leverages the non-uniqueness of SVD to minimize the average adjusted phase angles across all the MZIs in an IPNN, improving both the power efficiency and robustness in IPNNs. Unlike conventional pruning, our method preserves the trained weights, thereby guaranteeing zero inferencing-accuracy loss.\vspace{-0.05in}


\section{Proposed IPNN Optimization Using Non-Uniqueness of SVD under Reflections}
\vspace{-0.02in}

\begin{figure}[t]
  \centering
  \subfigure[]{
\includegraphics[width=.69\textwidth]{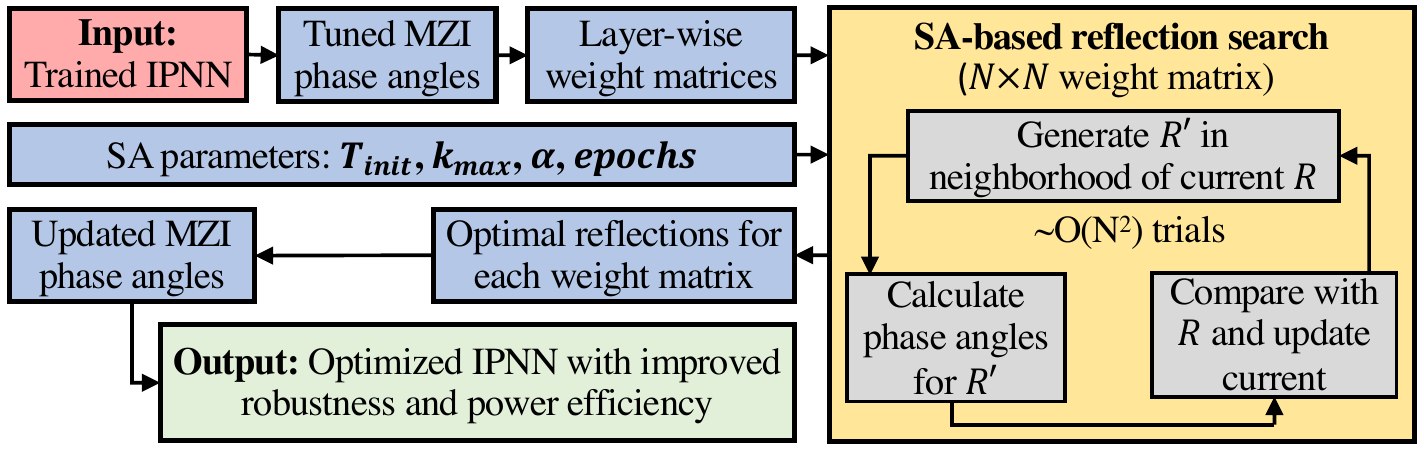}
}%
\subfigure[]{
\includegraphics[width=.29\textwidth]{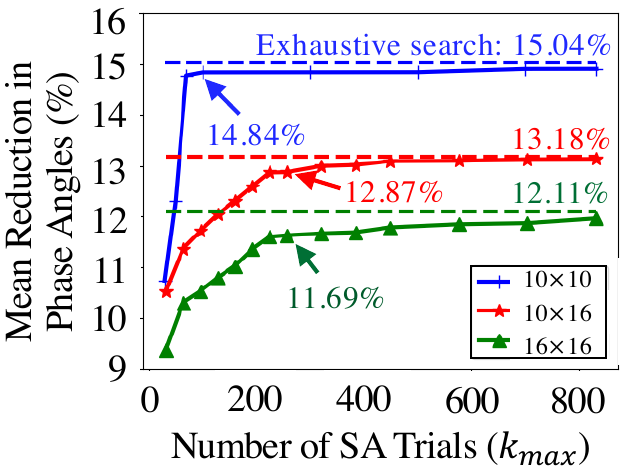}
}
\vspace{-1em}
  \caption{(a) Block diagram of the proposed SA-based optimal reflection search method. (b) Percentage reduction in the phase angles for five randomly generated 10$\times$10, 10$\times$16, and 16$\times$16 matrices with increasing number of trials. Dotted lines show the maximum possible reduction using an exhaustive search. Arrows show that the mean reduction saturates at around 256 trials for 10$\times$16 and 16$\times$16 matrices (100 trials for 10$\times$10 matrices).}
  \vspace{-2em}
  \label{fig2}
\end{figure}


Consider the weight matrix $M_{i}^{n_{i}\times n_{i-1}}$ of layer $L_{i-1}$ with $n_{i-1}$ neurons fully connected to the next layer $L_i$ with $n_i$ neurons in an IPNN. Using SVD, we have $M_{i}=U_{i}\Sigma_{i}V_{i}^H$, where $U_{i}$ and $V_{i}$ are unitary matrices with dimensions $n_{i}\times n_{i}$ and $n_{i-1}\times n_{i-1}$, respectively. Also, $V_{i}^H$ denotes the Hermitian transpose of $V_{i}$. 
Consider a \textit{reflector matrix} $R_i$ as a diagonal matrix with each element on the diagonal equal to $\pm 1$. It can be easily shown that for any non-singular diagonal matrix $D$, $R_iDR_i=D$ and $D^{-1}R_iD=R_i$. Therefore,  $M_i=U_{i}\Sigma_{i}V_{i}^H=U_{i}R_i\Sigma_{i}R_iV_{i}^H=U_{i}R_i\Sigma_{i}R_i^{H}V_{i}^H=U_{i}R_i\Sigma_{i}\left(V_{i}R_i\right)^H=U_{i}^{*}\Sigma_{i}V_{i}^{*H}$. Here, $U_{i}^{*}$ and $V_{i}^{*}$ denote the reflected forms of $U_i$ and $V_i$, respectively. Note that the same \textit{reflector} $R_i$ is applied to $U_i^{n_{i}\times n_{i}}$ and $V_i^{n_{i-1}\times n_{i-1}}$. Without loss of generality, we can assume that $n_i>n_{i-1}$. For each of the $2^{n_i}$ reflectors of $U_i$, the corresponding reflector of $V_i$ is given by its top-left $n_{i-1}\times n_{i-1}$ submatrix. The total number of reflections for $M_i^{n_i \times n_{i-1}}$ is therefore $2^{max(n_i, n_{i-1})}$.


In IPNNs, for any unitary matrix representation, the values of $\theta$ for all the MZIs remain constant across reflections, while $\phi$ for some MZIs can change by $\pm\pi$. 
The use of an appropriate reflector $R_i$ for each weight matrix minimizes the required phase angles, thus improving the IPNN power efficiency and robustness (see our discussion in Section 2). However, finding the optimal reflection for a given weight matrix using a naive brute-force search is computationally expensive as the number of possible reflections for $M_i^{n_i \times n_{i-1}}$ is $2^{max(n_i, n_{i-1})}$. To address this, we propose a heuristic based on simulated annealing (SA) to find an optimal reflection. For each weight matrix, we begin with a randomly selected reflector $R$. A new reflector $R'$ is then selected in the neighborhood of $R$. We define two reflectors to be \textit{neighbors} if they differ in only one diagonal element. The new reflector is accepted if it reduces the sum of the phase angles (\textit{downhill move}). \textit{Uphill moves} are also accepted with a probability of $e^{-\Delta/T}$ to prevent entrapment in a local minimum. Here, $\Delta$ and $T$ denote the change in the sum of phase angles and the annealing temperature, respectively. This process is repeated for \textit{epoch} times after which $T$ is reduced, i.e., $T_{new}=\alpha\cdot T_{old}$. We terminate SA when the total number of trials---which equals the number of new reflectors $R'$ generated---reaches a predetermined limit, $k_{max}$. Fig.~\ref{fig2}(a) shows a block diagram of the proposed SA-based search method. Simulation results show that our method can identify an optimal reflector for $M_{i}^{n_i\times n_{i-1}}$ in $O\left(max\left(n_i, n_{i-1}\right)^2\right)$ trials. \vspace{-0.05in}

\begin{figure}[t]
  \centering
  \hspace{-0.75em}
  \subfigure[10$\times$16 test matrices]{
\includegraphics[width=.247\textwidth]{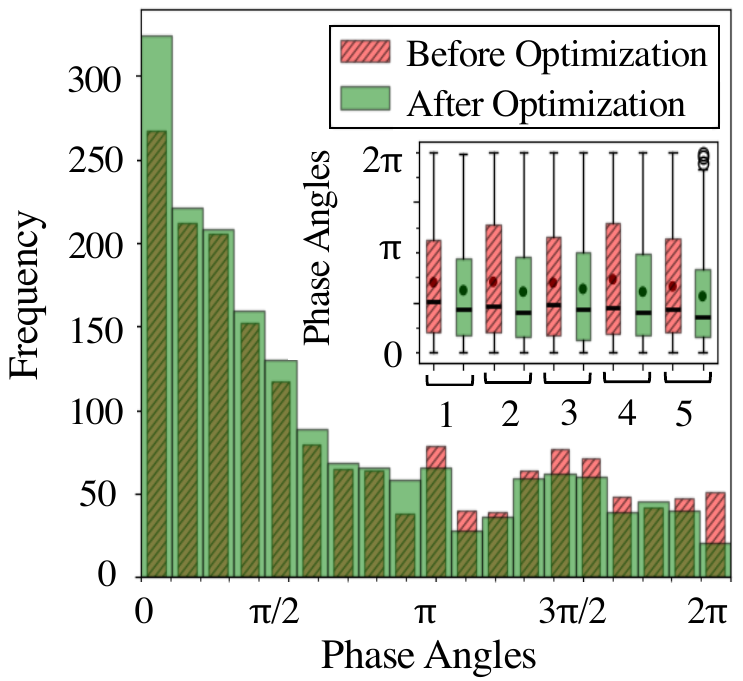}
}%
\hspace{-0.75em}
\subfigure[16$\times$16 test matrices]{
\includegraphics[width=.227\textwidth]{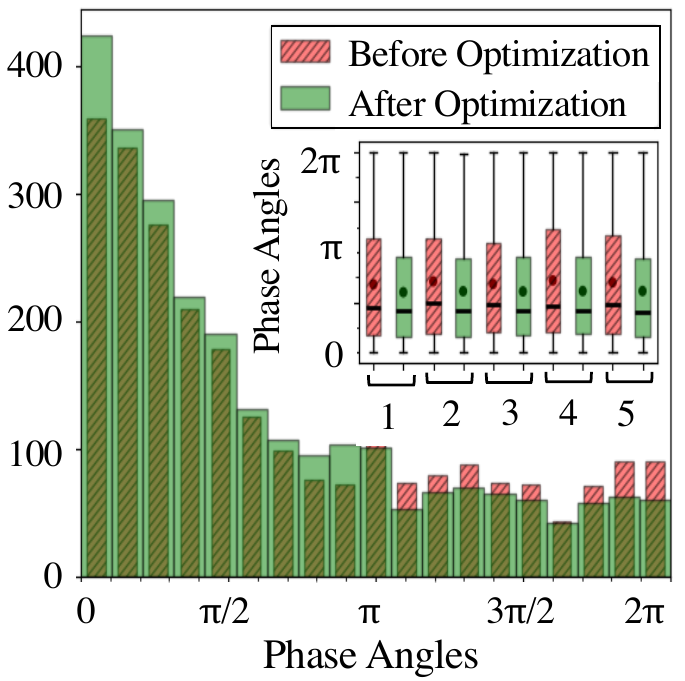}
}%
\hspace{-0.75em}
\subfigure[Trained IPNN]{
\includegraphics[width=.227\textwidth]{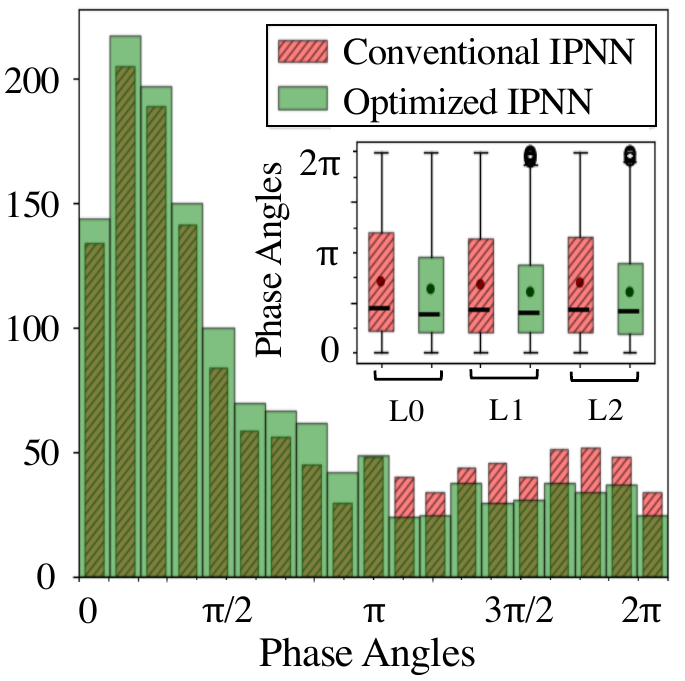}
}%
 \subfigure[Inferencing-accuracy loss]{
\includegraphics[width=.275\textwidth]{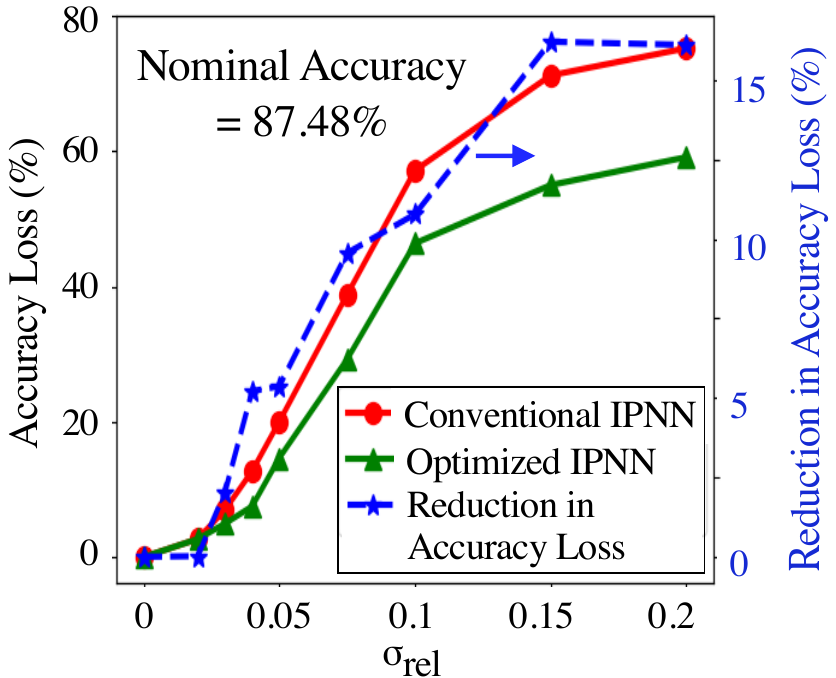}
}
\vspace{-1em}
  \caption{Histogram of the phase angles in (a) five randomly generated 10$\times$16 test matrices, and (b) five randomly generated 16$\times$16 test matrices, before and after optimization. Inset shows the box-plot with median and average (circles) of the phase angles in each test matrix (x-axis). (c) Histogram of the phase angles in the IPNN before and after optimization. Inset shows the box-plot of the phase angles in each layer (L0, L1, and L2 on the x-axis). (d) Mean accuracy loss under random uncertainties over 10 variation scenarios in the conventional and optimized IPNNs. Uncertainties in the phase angles are sampled from a Gaussian with their nominal value as the mean ($\mu$) and a standard deviation $\sigma=\sigma_{rel}\cdot \mu$. The blue dashed line shows the reduction in accuracy loss after optimization.}
  \vspace{-0.3in}
  \label{fig3}
\end{figure}

\section{Results and Discussion}
\vspace{-0.03in}
We show the efficiency of our proposed optimization method using the IPNN described in Section 2. The dimensions of the weight matrices in the IPNN are 16$\times$16 (input and first hidden layer) and 10$\times$16 (second hidden layer). To find the appropriate values of $T_{init}$, $\alpha$, and \textit{epoch} (see Fig.~\ref{fig2}(a)), we use a simplex-based parameter search method \cite{caserta2009metaheuristics} on five randomly generated 10$\times$10, 10$\times$16, and 16$\times$16 test matrices. To avoid bias, the trained weights for the MNIST classification task are not used for parameter search. While the optimum parameter values varied for the 10 test matrices, we found that the optimum values of $T_{init}$, $\alpha$, and \textit{epoch} were centered around 10, 0.8, and 2, respectively. Therefore, we considered these as the SA parameter values for subsequent simulations. A similar parameter search for five 10$\times$10 matrices generated the optimum values of $T_{init}$, $\alpha$, and \textit{epoch} to be 6.5, 0.8, and 2, respectively. Fig~\ref{fig2}(b) shows the mean reduction in the sum of the phase angles for the five test matrices (see above) of each dimension as the number of SA trials ($k_{max}$) increases. For each test matrix, we consider the average reduction over 10 rounds of optimization to ensure that any bias due to the randomly selected initial state in SA is canceled. The dotted lines show the maximum possible mean reduction obtained using an exhaustive search over the possible $2^{16}$ ($2^{10}$ for 10$\times$10 matrices) reflectors. We observe that the mean reduction with increasing $k_{max}$ saturates around $k_{max}=$~256 for 10$\times$16 and 16$\times$16 matrices ($k_{max}=$~100 for 10$\times$10); therefore, while an exhaustive search for the optimal reflection of an $N\times N$ matrix involves up to $2^N$ trials, our SA-based search requires only $O\left(N^2\right)$ trials. The histograms of the phase angles in the five 10$\times$16 and 16$\times$16 matrices (Figs.~\ref{fig3}(a) and \ref{fig3}(b), respectively) show that our optimization results in an increase (decrease) in the count of lower (higher) adjusted phase angles. This reduction in phase angles is achieved without affecting the matrix represented by the MZI network. Fig.~\ref{fig3}(c) shows the impact of our proposed optimization on the trained IPNN. Our method leads to an average reduction of 12.8\% in the overall network phase angles and consequently, the tuning-power consumption in the IPNN (with up to 15.3\% reduction in the phase angles in one layer). Fig. \ref{fig3}(d) compares the accuracy loss when random uncertainties are introduced in the conventional (not optimized) and optimized IPNN. The optimized IPNN leads to a lower accuracy loss and the improvement in robustness even increases at higher levels of uncertainties: e.g., at $\sigma_{rel}=~$0.2, our optimization reduces the accuracy loss by 16.1\%. Such improvements are significant as our optimization can be performed offline without impacting IPNN inferencing-accuracy.\vspace{-0.05in}

\section{Conclusions}
We have demonstrated a novel optimization method to improve power efficiency and robustness in coherent integrated photonic neural networks (IPNNs). Our method enhances IPNN power efficiency and robustness by reducing the susceptibility of MZIs with high adjusted phase angles to random uncertainties (due to inevitable fabrication-process variations and thermal crosstalk). It can be easily combined with bias control and mitigation techniques in IPNNs to facilitate the application and improve the efficiency of such techniques at no cost.\vspace{-0.05in}

\bibliographystyle{unsrt}
\bibliography{sample}

\end{document}